\documentclass[aps,prl,twocolumn]{revtex4}
\usepackage{graphicx}
\begin{document}

\title{Nonlinear diffusion model for Rayleigh-Taylor mixing}

\author{G. Boffetta$^{(1,2)}$, F. De Lillo$^{(1)}$, S. Musacchio$^{(3)}$}
\affiliation{$^{(1)}$Dipartimento di Fisica Generale and INFN, 
Universit\`a di Torino, via P.Giuria 1, 10125 Torino (Italy)  \\
$^{(2)}$ CNR-ISAC, Sezione di Torino, corso Fiume 4, 10133 Torino (Italy) \\
$^{(3)}$ CNRS, Lab. J.A. Dieudonn\'e UMR 6621,
Parc Valrose, 06108 Nice (France)}
\date{\today}

\begin{abstract}
The complex evolution of turbulent mixing in Rayleigh-Taylor 
convection is studied in terms of eddy diffusiviy models for 
the mean temperature profile. 
It is found that a non-linear model, derived within the general 
framework of Prandtl mixing theory, reproduces accurately the 
evolution of turbulent profiles obtained from numerical simulations. 
Our model allows to give very precise predictions for the turbulent heat
flux and for the Nusselt number in the ultimate state regime
of thermal convection.
\end{abstract}

\maketitle

Turbulent thermal convection is one of the most important manifestations
of turbulence. It appears in many natural phenomena, 
from heat transport in stars to atmosphere and oceanic mixing,
and it also plays a fundamental role in many technological 
applications \cite{siggia_arfm94}.

This Letter is devoted to the study of turbulent convection in the
Rayleigh-Taylor (RT) setup, a paradigmatic configuration in which a 
heavy layer of fluid is placed on the top of a light layer.
Gravitational instability at the interface of the two layers 
leads to a turbulent mixing zone which grows in time at the expenses 
of available potential energy \cite{sharp_physd84}.
Specific applications of RT convection range from cloud 
formation \cite{mammatus06}, 
to supernova explosion \cite{zwrdb_aj05,cc_natphys06} and
solar corona heating \cite{imsy_nat05}.
Because of the absence of boundaries, the phenomenology of RT turbulence
results simpler than other convective systems where the thermal
forcing is provided by walls, such as the Rayleigh-Benard configuration.

Recent theoretical work \cite{chertkov_prl03},
confirmed by numerical simulations
\cite{dly_jfm99,cc_natphys06,cabot_pof06,vc_pof09,bmmv_pre09,cmv_prl06,matsumoto_pre09}, 
predicts
for RT turbulence at small scales a turbulent cascade with 
Kolmogorov-Obukhov scaling (Bolgiano scaling in two dimensions).
Here we concentrate on large scale features of RT turbulence. 
We propose a simple closure scheme based on the general
framework of Prandtl mixing length theory and leading to a 
nonlinear diffusion model for temperature concentration.
Our closure reproduces with high accuracy the spatial-temporal
evolution of the mean temperature profile and
allows to derive a prediction for the scaling law of $Nu$ versus $Ra$
which fits perfectly data obtained from direct numerical simulations.


The equation of motion for the incompressible 
velocity field ${\bf v}$ (${\bf \nabla} \cdot {\bf v}=0$) 
and temperature field $T$ in the Boussinesq approximation is
\begin{eqnarray}
&& \partial_t {\bf v} + {\bf v} \cdot {\bf \nabla} {\bf v} = - {\bf \nabla} p
+ \nu \nabla^2 {\bf v} - \beta {\bf g} T \label{eq:1} \\
&& \partial_t T + {\bf v} \cdot {\bf \nabla} T = \kappa \nabla^2 T
\label{eq:2}
\end{eqnarray}
where $\beta$ is the thermal expansion coefficient,
$\nu$ the kinematic viscosity, $\kappa$ the thermal
diffusivity and ${\bf g}=(0,0,-g)$ is the gravitational acceleration.

\begin{figure}[htb!]
\includegraphics[width=7cm]{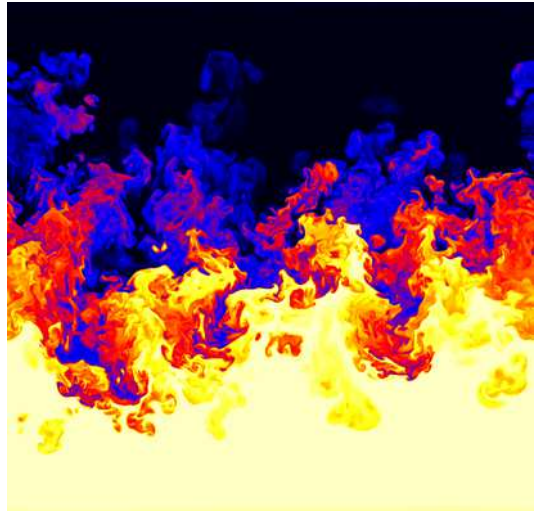}
\caption{(color online). 
Snapshot of a $(x,z)$ section of the temperature field 
for Rayleigh-Taylor turbulence numerical simulation. White (black)
represents hot, light (cold, heavy) fluid. 
Boussinesq equations (\ref{eq:1}-\ref{eq:2}) are integrated by a 
standard fully dealiased
pseudo-spectral code at resolution $N_x \times N_y \times N_z$
with $N_y=N_x$ and aspect ratio $L_x/L_z=N_x/N_z=r$ (here $N_x=1024$
and $r=1$). Other parameters are 
$\beta g=0.5$, $\theta_0=1$ ($A g=0.25$), $Pr=\nu/\kappa=1$ and
$\nu$ is chosen such that $k_{\max} \eta \ge 1.2$ in all runs
at final time. Initial perturbation is seeded by adding a 
$10\%$ of white noise to the initial temperature profile in a small
layer around $z=0$.
}
\label{fig1}
\end{figure}

The initial condition (at $t=0$) is a layer of cooler (heavier)
fluid on the top of a hotter (lighter) layer at rest,
i.e. ${\bf v}({\bf x},0)=0$ and 
$T({\bf x},0)=-(\theta_0/2) \mbox{sgn}(z)$ where $\theta_0$
is the initial temperature jump which fixes the
Atwood number $A=(1/2) \beta \theta_0$ ($T=0$ is the reference 
mean temperature).
This configuration is unstable and after the linear instability phase,
the system develops a turbulent mixing zone which grows in time
starting from the plane $z=0$. An example of the turbulent temperature
field obtained from high resolution direct numerical simulations of 
(\ref{eq:1}-\ref{eq:2})
is shown in Fig.~\ref{fig1}.

In the mixing layer turbulent kinetic energy $E=(1/2) \langle v^2 \rangle$
is produced at the expense of potential energy $P=-\beta g \langle z T \rangle$
as the energy balance indicates
\begin{equation}
- {d P \over dt} = \beta g \langle w T \rangle = 
{d E \over dt} + \varepsilon_{\nu}
\label{eq:3}
\end{equation}
where $\varepsilon_{\nu}=\nu \langle (\partial_{\alpha} v_{\beta})^2 \rangle$
is the viscous energy dissipation and $\langle \rangle$ represents the
integral over the physical domain.
Assuming that in the turbulent state all quantities in (\ref{eq:3}) 
scale in the same way one can 
balance $d v_{rms}^2/dt \simeq \beta g \theta_0 v_{rms}$ 
(because temperature fluctuations are bounded by the initial jump $\theta_0$) 
and therefore one obtains the temporal scaling of velocity fluctuations
$v_{rms} \simeq \beta g \theta_0 t \simeq A g t$,
i.e. a motion forced with constant acceleration $g$. 

The accelerated growth of the width of the mixing layer 
is one of the standard diagnostics in the studies of RT turbulence
\cite{dly_jfm99,clark_pof03,hdabrc_prl07,krbghca_pnas07}.
Several definitions
for the width have been proposed, based on either local or global
properties of the mean temperature profile
$\overline{T}(z,t) \equiv 1/(L_x L_y) \int T({\bf x},t)dx dy$.
The simplest measure $h_r$ is based on the threshold value of $z$ at which 
$\overline{T}(z,t)$ reaches a fraction $r$ of the maximum value i.e.
$\overline{T}(\pm h_r(t)/2,t)= \mp r \theta_0/2$ \cite{dly_jfm99}.
This local definition of $h$ can be rather noisy and therefore
alternative definitions based on integral quantities 
have been proposed \cite{as_pof90,dly_jfm99,cc_natphys06}
\begin{equation}
h_M \equiv \int_{-L_z/2}^{L_z/2} M(\overline{c}) dz
\label{eq:4}
\end{equation}
where $c=(T_{max}-T)/(T_{max}-T_{min})=1/2-T/\theta_0$ is the
normalized dimensionless temperature ($0 \le c \le 1$)
and $M$ is a mixing function which has support on the mixing
layer only, e.g. a logistic function $M(c)=4c(1-c)$ \cite{vc_pof09} or
a tent function $M(c)=2 c + (2-4c) \theta(c-1/2)$ \cite{cc_natphys06}.
Dimensionally, $h$ is expected to grow with accelerated law
$h(t)=\alpha A g t^2$ with the dimensionless coefficient $\alpha$ which
depends on the definition of $h$ and apparently also on the 
form of the initial perturbation of the interface 
\cite{dimonte_etal_pof04,krbghca_pnas07}.
Recent studies \cite{rc_jfm04,cc_natphys06} have shown that a
more robust and consistent determination of $\alpha$ can be obtained if an
initial time $t_0 \ne 0$ is taken into account (physically
representing the offset at which the $t^2$ law sets in) suggesting the 
possibility of a universal value, independent on the form of the initial
perturbation.

\begin{figure}[htb!]
\includegraphics[width=8cm]{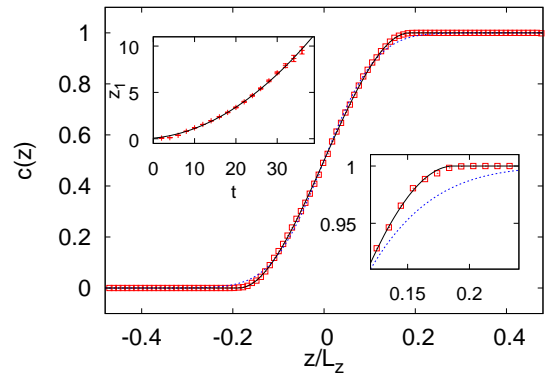}
\caption{(color online). Normalized mean temperature profile $c(z)$
computed by averaging over horizontal planes 
the turbulent temperature field of Fig.~\ref{fig1}.
Blue dotted line is the prediction of the linear diffusion
model (\ref{eq:7}). Black continuous line is the fit with the nonlinear 
model (\ref{eq:10}). 
Lower inset: enlargement of the temperature profile at the edge of 
the mixing layer. 
Upper inset: evolution of $z_1$ obtained by fitting the 
temperature profile with (\ref{eq:10}) 
at different times and over four different realizations.
The line represents the fit $z_1=\gamma A g (t+t_0)^2$ which gives 
$\gamma \simeq 0.025$ and $t_0 \simeq 3.3$.
}
\label{fig2}
\end{figure}

The evolution equation for the normalized temperature profile 
$\overline{c}(z,t)$ is obtained 
by averaging (\ref{eq:2}) over the horizontal directions (assumed
periodic)
\begin{equation}
\partial_t \overline{c} + \partial_z \overline{w c} = 
\kappa \partial_z^2 \overline{c}
\label{eq:5}
\end{equation}
where $w$ represent the vertical velocity. The thermal flux term
$\overline{w c}$ makes (\ref{eq:5}) not closed. Following a common 
approach in turbulence, we close this equation in terms of 
an eddy diffusivity $K(z,t)$ so that (\ref{eq:5}) is rewritten as
\begin{equation}
\partial_t \overline{c} = 
\partial_z K(z,t) \partial_z \overline{c}
\label{eq:6}
\end{equation}
Molecular diffusivity $\kappa$, included additively in $K(z,t)$,
can be neglected for large scale properties at high P\'eclet number.
The simplest approximation is to consider $K$ independent on $z$. 
For our problem, being a diffusion coefficient (i.e. a velocity time 
a scale) the eddy diffusivity is expected to depend on $t$ as
$K(t)=b^2 (A g)^2 t^3$ with $b$ a free dimensionless parameter.
The self-similar solution to (\ref{eq:6})
with a step initial condition $\overline{c}(z,0)=\theta(z)$ 
is 
\begin{equation}
\overline{c}(z,t)= {1 \over 2} \left[ 1+ \mbox{erf} 
\left({z \over b A g t^2} \right) \right]
\label{eq:7}
\end{equation}

\begin{figure}[htb!]
\includegraphics[width=8cm]{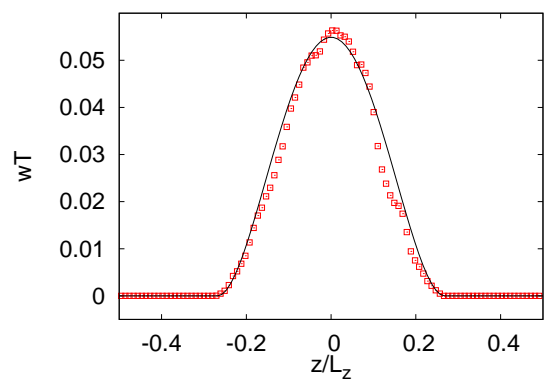}
\caption{(color online). Heat flux profile $\overline{w T}$ obtained 
at the same time of Fig.~\ref{fig2}. 
Black line represents the prediction of the nonlinear diffusion model
as discussed in the text.}
\label{fig3}
\end{figure}

The constant diffusivity solution (\ref{eq:7}) is a 
relatively good approximation 
of the actual profile obtained from the numerical simulations
of the full set of equations (\ref{eq:1}-\ref{eq:2}), as shown in 
Fig.~\ref{fig2}.
A closer inspection of the figure reveals that the model profile 
(\ref{eq:7}) is smoother than the actual profile at the edges of 
the mixing region (see inset of Fig.~\ref{fig2}).
The physical origin of this discrepancy is that turbulent mixing is
not homogeneous within the mixing layer. Indeed turbulent velocity 
fluctuations decrease at the ends
of the mixing region, and therefore a constant $K$ overestimates
the diffusivity in these regions.

An improved model must therefore 
take into account a $z$-dependence of the diffusivity. 
Within the general framework of 
mixing length theory by Prandtl \cite{prandtl25,siggia_arfm94}, 
the eddy diffusivity can be written as
$K(z,t) \simeq H^2 \partial_z V$ where $H$ represents a 
length characteristic of mixing and $V$ is the typical velocity
fluctuation. 
Because velocity is driven by buoyancy at large scale, from
equation (\ref{eq:1}) one can estimate that
after a time $t$ the typical velocity is 
$V \propto \beta g T t$ and taking $H \propto h(t)$
one obtains for the eddy diffusivity 
$K(z,t) = a (A g)^3 t^5 \partial_z \overline{c}$ where $a$ is again
a dimensionless constant to be determined empirically.
We remark that a similar approach, based on 
gradient dependent diffusivity, has been 
recently used for successfully modeling mixing 
in stratified flows \cite{ocre_prl09}.
Inserting the above expression in (\ref{eq:6}) 
one obtains a nonlinear diffusive model for the mean temperature
profile 
\begin{equation}
\partial_t \overline{c} = 
a (A g)^3 t^5 \partial_z (\partial_z \overline{c})^2
\label{eq:8}
\end{equation}
Observe that the non-linearity of (\ref{eq:8}) reflects the 
fact that temperature fluctuations are not passive in this problem
as they drive velocity fluctuations in (\ref{eq:1}).

Introducing the concentration derivative 
$\varphi(z,t)=3/(a (Ag)^3) \partial_z \overline{c}(z,t)$
and a new time variable $t'=t^6$, (\ref{eq:8}) is rewritten in a more 
standard form
\begin{equation}
\partial_{t'} \varphi = \partial_z (\varphi^n \partial_z \varphi)
\label{eq:9}
\end{equation}
with $n=1$. 
Equation (\ref{eq:9}) represents a class of nonlinear diffusion equations
with concentration dependent diffusivity well studied in different 
fields such as 
thermal waves in plasma radiation \cite{zr_arfm69} and
diffusion problems in porous media where for 
our case $n=1$ equation (\ref{eq:9}) is also known with the name of 
Boussinesq equation \cite{bear88}.
The value of $n$ governs 
the behavior of the gradient when $\varphi \to 0$ which is finite for 
the present case.
The self-similar solution (for general $n$ and dimensionality) is
known \cite{pattle59} and gives for our case
\begin{equation}
\begin{array}{lr}
\overline{c}(z,t)={1 \over 4} {z \over z_1} \left[3 -
\left({z \over z_1}\right)^2 \right] + {1 \over 2} & \qquad |z| \le z_1 \\
\overline{c}(z,t)= 0 & z < -z_1 \\
\overline{c}(z,t)= 1 & z > z_1
\end{array}
\label{eq:10}
\end{equation}
where $z_1(t)=\gamma A g t^2$ with $\gamma=(3 a/2)^{1/3}$.

Having the analytical expression (\ref{eq:10}) for the mean concentration,
the different definitions of the width of the mixing layer are
all expressed in terms of $z_1$ and differ by a factor only
(e.g.  $h_1=2 z_1$ and $h_M=(3/4) z_1$ for the tent function 
\cite{cc_natphys06}).
Figure~\ref{fig2} shows that the polynomial function (\ref{eq:10})
fits very well the mean concentration profile obtained from numerical
simulations. Runs at different resolutions (and viscosity, the only
parameter in (\ref{eq:1}-\ref{eq:2}) when $Pr=1$) give analogous
results. 
By fitting the numerical profiles at different times, one obtains the
evolution of $z_1$ displayed in the inset of Fig.~\ref{fig2}
which is consistent with the quadratic law
$z_1=\gamma A g (t+t_0)^2$ ($t_0$ is the reference time as discussed
above). The value obtained in this way for the coefficient is 
$\gamma=0.025 \pm 0.002$ which for the profile $h_M$ gives
$\alpha=(3/4) \gamma \simeq 0.019$ in agreement with previous 
numerical results \cite{cc_natphys06,vc_pof09}.

The nonlinear diffusion model can be extended from 
geometrical quantities to study the evolution of dynamical 
properties of turbulent convection. 
In particular, in the limit of small thermal
diffusivity, from (\ref{eq:5}) and (\ref{eq:8}) one has an
expression for the turbulent heat flux in terms of the mean
temperature profile $\overline{w T}=a (Ag)^3 t^5 (\partial_z T)^2/\theta_0$.
Figure~\ref{fig3} shows that the numerically measured profile 
of the heat flux is indeed quite close to the model prediction, 
a justification {\it a posteriori} of the proposed nonlinear 
closure scheme.
Using the definition in (\ref{eq:3}) the loss of 
potential energy in kinetic energy (and dissipation) is written as
$-dP/dt=(4/5) \gamma^2 (Ag)^3 t^3$ which shows that $\gamma$ is
a measure of the efficiency of conversion of available potential
energy in the turbulent flow.

\begin{figure}[htb!]
\includegraphics[width=8cm]{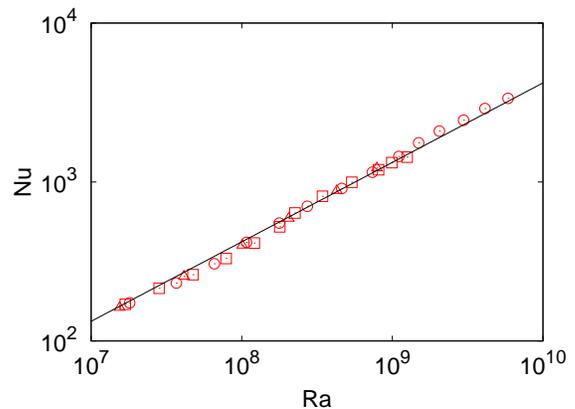}
\caption{Nusselt number $Nu=\langle w T \rangle/(\kappa \theta_0)$
versus Rayleigh number $Ra=A g h^3/(\nu \kappa)$ from three different set
of simulations at resolutions $256 \times 256 \times 1024$ 
(squares), $512 \times 512 \times 2048$ 
(circles) and $1024 \times 1024 \times 1024$ (triangles) at $Pr=1$. Kinematic 
viscosities for the three runs are respectively $\nu=6 \times 10^{-4}$, 
$\nu=3 \times 10^{-4}$ and $\nu=1 \times 10^{-4}$.
The line is the prediction (\ref{eq:11}) with $\gamma=0.025$.
}
\label{fig4}
\end{figure}

The relation between the heat flux and the profile geometry
can be reformulated in terms of dimensionless quantities.
Indeed, integrating over the width of the mixing layer 
it gives a relation between the Nusselt number 
$Nu=\langle w T \rangle/(\kappa \theta_0)$
(the ratio of convective to conductive heat transfer)
and the Rayleigh number $Ra=A g h^3/(\nu \kappa)$
(the ratio of the buoyancy forces to diffusivities).
Using the expression (\ref{eq:10}) and for the length 
$h=h_1=2 z_1$, one obtains the temporal evolution laws for
the two quantities as
$Ra=8 \gamma^3 (A g)^4 t^6/(\nu \kappa)$, 
$Nu=2 \gamma^2 (A g)^2 t^3/(5 \kappa)$ and therefore 
the relation
\begin{equation}
Nu = {1 \over 5 \sqrt{2}} \gamma^{1/2} Pr^{1/2} Ra^{1/2}
\label{eq:11}
\end{equation}

Equation (\ref{eq:11}) represents the well known Kraichnan's
prediction for the ``ultimate state of thermal convection'' 
\cite{kraichnan_pof62,gl_jfm00}
which is a regime of turbulent convection expected to hold
when the contribution of thermal and kinetic boundary layers 
becomes negligible.
Because of the absence of boundaries, 
RT turbulence is a natural candidate 
for the appearance of this regime which has indeed been observed 
recently in numerical simulations both in two and three dimensions
\cite{lt_prl03,cmv_prl06,bmmv_pre09}. 
Figure~\ref{fig4} shows that the prediction (\ref{eq:11})
with $\gamma=0.025$ fits well the numerical data obtained from a 
set of simulations at different resolutions. 
The fact that $Nu \gg 1$ is a posteriori confirmation of the negligible
contribution of thermal diffusivity.

It is interesting to observe that the above result for $Nu$ 
satisfies a general bound which can be easily obtained starting
from (\ref{eq:5}). Neglecting thermal diffusivity and assuming
a self-similar evolution of the profile $\overline{c}(z,t)=f(z/z_1(t))$
with the symmetry condition $f(-z)=1-f(z)$, integrating (\ref{eq:5})
over the $z$ domain $[-L_z/2,L_z/2]$ twice, one obtains
\begin{equation}
\int_{-L_z \over 2}^{L_z \over 2} dz \overline{w c} = {2 \dot{z}_1 \over z_1} 
\left[ 2 \int_0^{L_z \over 2} dz z \overline{c}(z,t) - {L_z^2 \over 4} \right]
\label{eq:12}
\end{equation}
Using the fact that for $z>0$ $\overline{c}(z,t)>1/2$ and assuming that the 
flow is still unmixed, $\overline{c}(z,t)=1$ for $z>z_1$, we get a bound
\begin{equation}
Nu = - {1 \over \kappa} \int_{-L_z \over 2}^{L_z \over 2} dz \overline{w c}
\le {1 \over \kappa} z_1 \dot{z}_1
\label{eq:13}
\end{equation}
If we now further assume the accelerated growth of the mixing layer,
$z_1(t)=\gamma A g t^2$ we end with a bound on the growth 
of the Nusselt number
\begin{equation}
Nu \le {2 \over \kappa} \gamma^2 (A g)^2 t^3
\label{eq:14}
\end{equation}
which is indeed satisfied by our model.
The physical interpretation of this bound is transparent: the growth
of the heat flux follows the dimensional $t^3$ law with a coefficient
which depends on the shape of the mean temperature profile. 
Maximum growth (\ref{eq:14}) is achieved when $c(z,t)=1/2$ 
for $-z_1 \le z \le z_1$ which means a perfect mixing within
the mixing layer. This would correspond to a coefficient $(\gamma/2)^{1/2}$
in (\ref{eq:11}).

In this Letter we have introduced a nonlinear diffusion
model with a gradient dependent eddy diffusivity 
which reproduces accurately the large scale phenomenology
of Rayleigh-Taylor turbulence obtained from high-resolution
numerical simulations. 
The model contains a single free 
parameter, a measure of the turbulence production efficiency,
which is directly related to the rate of accelerated growth 
of the mixing layer.
The proposed closure scheme represents 
an important step for a phenomenological description of RT
turbulence as it connects the evolution of the Nusselt number
to the growth of the mixing layer, a global 
geometrical quantity which can be easily obtained in experiments.


\bibliography{biblio}{}

\end{document}